\documentclass[11pt,a4paper]{article}
\pdfoutput=1
\usepackage{jheppub}

\usepackage{xspace}

\newcommand{\bea}{\begin{eqnarray}}
\newcommand{\eea}{\end{eqnarray}}
\newcommand{\beq}{\begin{equation}}
\newcommand{\eeq}{\end{equation}}

\newcommand{\tb}{\ensuremath{\tan\beta}\xspace}

\newcommand{\Btn}{\ensuremath{\mathrm{BR}(B^+ \to\tau^+\nu_\tau)}\xspace}
\newcommand{\bsg}{\ensuremath{b \to s\gamma}\xspace}

\def\stilde{\widetilde}

\newcommand{\Neu}[1]{\ensuremath{\stilde \chi_{#1}^0}\xspace}

\def\vev{\emph{vev}\xspace}

\title{Higgs and non-universal gaugino masses: no SUSY signal expected yet?}
\author[a,b]{Sascha Caron,}
\author[a,b]{Jari Laamanen,}
\author[a,b]{Irene Niessen}
\author[c]{Antonia Str\"ubig}
\affiliation[a]{Experimental and Theoretical High Energy Physics,
  IMAPP, Faculty of Science,  Radboud University Nijmegen,
  Mailbox 79,
  P.O. Box 9010,
  NL-6500 GL Nijmegen, The Netherlands}
\affiliation[b]{Nikhef, Science Park 105, 1098 XG Amsterdam, The Netherlands}
\affiliation[c]{Physikalisches Institut, University of Freiburg, Hermann-Herder Str.3, 79104 Freiburg, Germany}
  
\emailAdd{Sascha.Caron@cern.ch}
\emailAdd{j.laamanen@science.ru.nl}
\emailAdd{i.niessen@science.ru.nl}
\emailAdd{antonia.struebig@cern.ch}

\abstract{
So far, no supersymmetric particles have been detected at the
Large Hadron Collider (LHC). However, the recent Higgs results have interesting implications for the SUSY parameter space. 
In this paper, we study the consequences of an LHC Higgs signal for a model with
non-universal gaugino masses in the context of SU(5) unification.
The gaugino mass ratios associated with the higher representations
produce viable spectra that are largely inaccessible to the current
LHC and direct dark matter detection experiments.
Thus, in light of the Higgs results, the non-observation of SUSY is no surprise.
}


\begin{document}

\maketitle

\section{Introduction}
\label{sec:intro}

Broken low-scale supersymmetry (SUSY) is one of the most relevant
candidates of new physics. It may explain the hierarchy problem
\cite{Witten:1981nf,Dimopoulos:1981zb,Sakai:1981gr,Kaul:1981hi} and
facilitates a unification of the running gauge couplings at the grand
unification (GUT) scale
\cite{Ellis:1990wk,Giunti:1991ta,Amaldi:1991cn,Langacker:1991an}.
In the minimal supersymmetric standard model (MSSM) with $R$-parity
\cite{Salam:1974xa, Fayet:1974pd,Farrar:1978xj, Dimopoulos:1981dw,
  Farrar:1982te} imposed, it can also provide an appropriate dark
matter (DM) particle.  At the same time, SUSY has to reproduce known
physics at lower energies. In this light, the recent signs of a
possible Higgs boson at the Large Hadron Collider (LHC)
\cite{Collaboration:2012si,Chatrchyan:2012tx}
can put tight constraints on supersymmetry.

The number of free parameters in the MSSM is notoriously large,
leading to a loss of predictability. Therefore, the MSSM parameter
space is usually restricted by a combination of theoretical and
phenomenological considerations. This has lead to a study of
simplified models such as the constrained MSSM (CMSSM), which has a
universal scalar mass $m_0$ and trilinear parameter $A_0$ at the GUT
scale as well as universal gaugino masses. However, why would the ratios of the different gaugino masses be
unity at the GUT scale? A close inspection of the structure of the SUSY breaking terms in fact shows that different ratios arise quite naturally. As an example we will
consider a model of non-universal gaugino masses
\cite{Ellis:1985jn,Drees:1985bx} in the context of SU(5)
unification. Note that the SU(5) group may be embedded into a larger
GUT group, like SO(10) or E(6) \cite{Martin:2009ad}, in which case the
studied gaugino mass ratios belong to the simplest branching rules of
these embeddings.

The effect of the usual low-energy and DM constraints on non-universal
gaugino mass models has been discussed in
\cite{Anderson:1996bg,Anderson:2000,Huitu:2000,
  Corsetti:2001, Birkedal-Hansen:2003, Huitu:2005wh, Belanger:2005,
  Ananthanarayan:2007fj, Bhattacharya:2007dr, Huitu:2008sa,
  Drees:2008tc, Bandyopadhyay:2008sd, Huitu:2009zs,
  Chattopadhyay:2009fr, Martin:2009ad, Bhattacharya:2009wv,
  Okada:2011wd, Bi:2011ha, Younkin:2012ui}. We will focus on the
impact of the recent LHC Higgs results and the effects of the ATLAS
squark and gluino searches. Finally, we will study how this affects direct DM detection.

\section{Model and Parameters}
\label{sec:su5}

Many supergravity-type models use unified gaugino masses at the GUT
scale, but there is no compelling reason to do so. In general, the
mass-generating terms for the gauginos $\lambda$  have the form
\cite{Cremmer:1982wb,Cremmer:1982en, Ellis:1985jn}:
\begin{equation}
{\cal L}_{\mathrm{gaugino\; mass}}\sim \langle F_{ab}\rangle\lambda^a\lambda^b+c.c.\,.\label{eq:lagrterm}
\end{equation}
The Lagrangian must be invariant under the gauge symmetry group, which
we take to be SU(5). Gauginos, like the corresponding gauge bosons,
reside in the adjoint representation of the gauge group, which in SU(5) is
24-dimensional. The gaugino product in Eq.~\eqref{eq:lagrterm}
transforms under SU(5) according to a
representation appearing in the symmetric product of two adjoint
representations: 
\bea ({\bf 24 \otimes 24})_{Symm} = {\bf 1 \oplus 24
  \oplus 75 \oplus 200}\,.
\label{product}
\eea %
The mass-generating term $F_{ab}$, which contains a vacuum expectation
value (\vev), must be in the same representation as the gaugino
product in order to make the Lagrangian invariant. Therefore, it can
be in any of the representations on the right-hand side of Eq.~\eqref{product}. The
representation $\bf1$ corresponds to the CMSSM, where the generated
gaugino masses are unified at the GUT scale. Any of the other
representations, however, will yield different mass relations at the
GUT scale
\cite{Ellis:1985jn,Drees:1985bx,Anderson:1996bg,Anderson:2000}. Such
different mass relations occur naturally in a GUT and have important
consequences for SUSY phenomenology.
Table~\ref{tab1} shows the ratios of resulting gaugino masses at
tree-level at the GUT scale and at one-loop level at the electroweak
(EW) scale. We study the case of each representation separately,
although an arbitrary combination of these is also allowed (see, e.g.,
\cite{Huitu:2008sa,Chattopadhyay:2009fr,Younkin:2012ui}).
\begin{table}[htb]
  \caption{\label{tab1} Ratios of the gaugino masses at the GUT scale
    in the normalization $M_3$ = 1, and at the EW
    scale in the normalization \mbox{$M_3^{EW}$ = 1} at the 1-loop level
    \cite{Huitu:2005wh}.}
  \centering
  \begin{tabular}{c|ccc|ccc}
    %\hline
    rep & $M_1$ & $M_2$ & $M_3$ & 
    $M_1^{EW}$ & $M_2^{EW}$ & $M_3^{EW}$
    \\  \hline {\bf 1} & 1 & 1
    & 1 & 0.14 & 0.29 & 1
    \\ {\bf 24} & -0.5 & -1.5 & 1 & -0.07 & -0.43 & 1
    \\ {\bf 75} & -5 & 3 & 1 & -0.72 & 0.87 & 1
    \\ {\bf 200} & 10 & 2 & 1 &1.44 & 0.58 & 1
    \\ %\hline 
  \end{tabular}
\end{table}

In our scans of the parameter space, we sample $M_3$ at the GUT
scale. This variable correlates with the gluino mass and fixes the
other gaugino masses for each representation according to
Table~\ref{tab1}. In addition, we vary the scalar mass $m_0$, the
trilinear parameter $A_0$, the ratio of the Higgs \vev's
$\tan\beta=\langle H_u^0\rangle/\langle H_d^0\rangle$, and the sign of
the SUSY Higgs mass parameter $\mu$. The parameter space is sampled
quasi-randomly \cite{niederreiter1988,niederreitercode} using a flat
prior in all variables. We have checked that the results are unchanged for a logarithmic prior and provide the relevant figures in a supplement. For each representation, we
sampled 150.000 points. The scan ranges are given in
Table~\ref{t:ranges}.
\begin{table}[htb]
  \caption{\label{t:ranges} Parameter ranges.}
  \centering
  \begin{tabular}{c|c}
%&flat prior\\ %&log prior\\
parameter & range \\
\hline
$m_0,M_3$ (GeV)&[100\,,\,3000]\\ %&[100\,,\,3000]\\
$A_0$ (TeV)&[-7\,,\,7]\\ % &[-7\,,\,7]\\
$\tan\beta$&[2.5\,,\,59]\\ %&[2.5\,,\,59]\\
$\mu$&$\pm1$ %&$\pm1$
  \end{tabular}
\end{table}
The particle spectrum was calculated using SOFTSUSY
(v.3.1.7)~\cite{Allanach:2001kg}. A top
pole mass of $m_t = 173.3$ GeV was used throughout this study. 

% some non-univ sacalars
% \cite{Chattopadhyay:2008hk, Bhattacharya:2009ij}}

\section{Constraints}
\label{sec:allconstraints}

After sampling the parameter space, we select viable models by
requiring they satisfy a number of constraints. We will briefly list
these constraints, which have all been implemented at 95\% CL using micrOmegas (v.2.4.1)~\cite{Belanger:2006is}.

Combining the present experimental value of the $B \to X_s \gamma$
branching ratio \cite{Asner:2010qj}
% \begin{eqnarray}
%  BR(B \to X_s \gamma) &=& (355 \pm 24 \pm
% 9) \times 10^{-6}.\nonumber
% \end{eqnarray}
% In our constraint, the 
with the theoretical uncertainties 
% are included as well: in the SM at the NNLO QCD level the
% uncertainty can be estimated to be $23 \times 10^{-6}$
% \cite{Misiak:2006zs}, in the MSSM, the theoretical uncertainty is
% estimated to be additional 5\% (we take $15 \times 10^{-6}$)
\cite{Misiak:2006zs,Ellis:2007fu}, gives
\cite{Huitu:2011cp}:
\begin{equation}
  \label{eq:4}
  BR(B \to X_s \gamma) = (355 \pm 142) \times 10^{-6}.
\end{equation}
The \bsg constraint is sensitive to the sign of $\mu$
\cite{Nath:1994tn}, preferring the positive value.

We also use the $B \to \tau\nu$ branching ratio as a constraint by demanding \cite{Isidori:2006pk,Bhattacherjee:2010ju}:
\begin{equation}
0.99<\frac{\mathrm{BR}(B^+\to\tau^+
    \nu_\tau) _{\mathrm{\,SUSY}} }
  {\mathrm{BR}(B^+\to\tau^+\nu_\tau)_{\mathrm{SM}}}< 3.19,  \label{RatioExp}
\end{equation}
where the numerator denotes the branching ratio in the SUSY scenario,
including the SM contribution.  The constraint \eqref{RatioExp} tends
to prefer small values of \tb in order not to decrease the ratio too
much below the lower limit.

Furthermore, we use a conservative upper limit of $BR(B_s \to \mu^+ \mu^-) < 5.0
\times 10^{-8}$. 

From the 7-year Wilkinson Microwave Anisotropy Probe (WMAP) results,
the cold DM relic density in the universe \cite{Komatsu:2010fb} with a
10\% theoretical uncertainty added \cite{Baro:2007em}, is given by
\begin{equation}
  \label{wmaplimits}
  0.0941
  < \Omega_c h^2 <
  0.131.
\end{equation}
Since DM may be of non-supersymmetric origin, we only impose the upper
limit of constraint \eqref{wmaplimits}. This severely reduces the
number of viable models for the $\bf 1$ and the $\bf 24$
representations, where the LSP is generally Bino-like. In contrast, in
the $\bf75$ representation the LSP can also be Higgsino-like, while in
the $\bf200$ representation, the LSP is either Wino- or Higgsino-like. For such
models, the annihilation cross section in the early universe is
higher, naturally leading to a lower DM relic density.
Whereas points
with a Bino-like LSP are heavily constrained by the DM relic density,
points with a Wino- or Higgsino-like LSP thus tend to survive the DM
constraint.

Because of the lack of consensus on how to treat the anomalous
magnetic moment of the muon \cite{Bodenstein:2011qy}, we have opted
not to include it as a constraint. We do, however, take into
account the SUSY particle mass limits obtained by LEP as implemented in
micrOmegas.

\section{Consequences of the Higgs Mass}
\label{sec:Higgs}

The recent LHC results have excluded a Higgs boson with a mass beyond
127 GeV. They do, however, show an excess in the mass range $121 - 127$~GeV. Encouraged by these results, we study the consequences of
a Higgs boson with a mass between $121$~GeV~$\leq m_H\leq127$~GeV. As this is quite a high mass for a SUSY Higgs, one would expect that this constraint limits the parameter space. In particular, such a high mass requires sizeable corrections from the third generation squarks, thus requiring them to be heavy. Therefore, we will first study the effect of this constraint on the stop masses. Secondly, we will discuss the consequences for squarks and gluinos, since these particles play a crucial role in the SUSY searches at the LHC.

In this section, we will show how the parameter space changes due to the Higgs mass constraint. In all plots, the
light grey boxes indicate the number of models in a given bin that
have a neutralino LSP, with a bigger box standing for a larger number of models. In the same way, the dark grey boxes indicate how many models
pass the low-energy and DM constraints listed in
section~\ref{sec:allconstraints}. The black points are models that
also have the correct value of the Higgs mass.

\begin{figure}[htb]
\includegraphics[height=0.49\linewidth,angle=90]{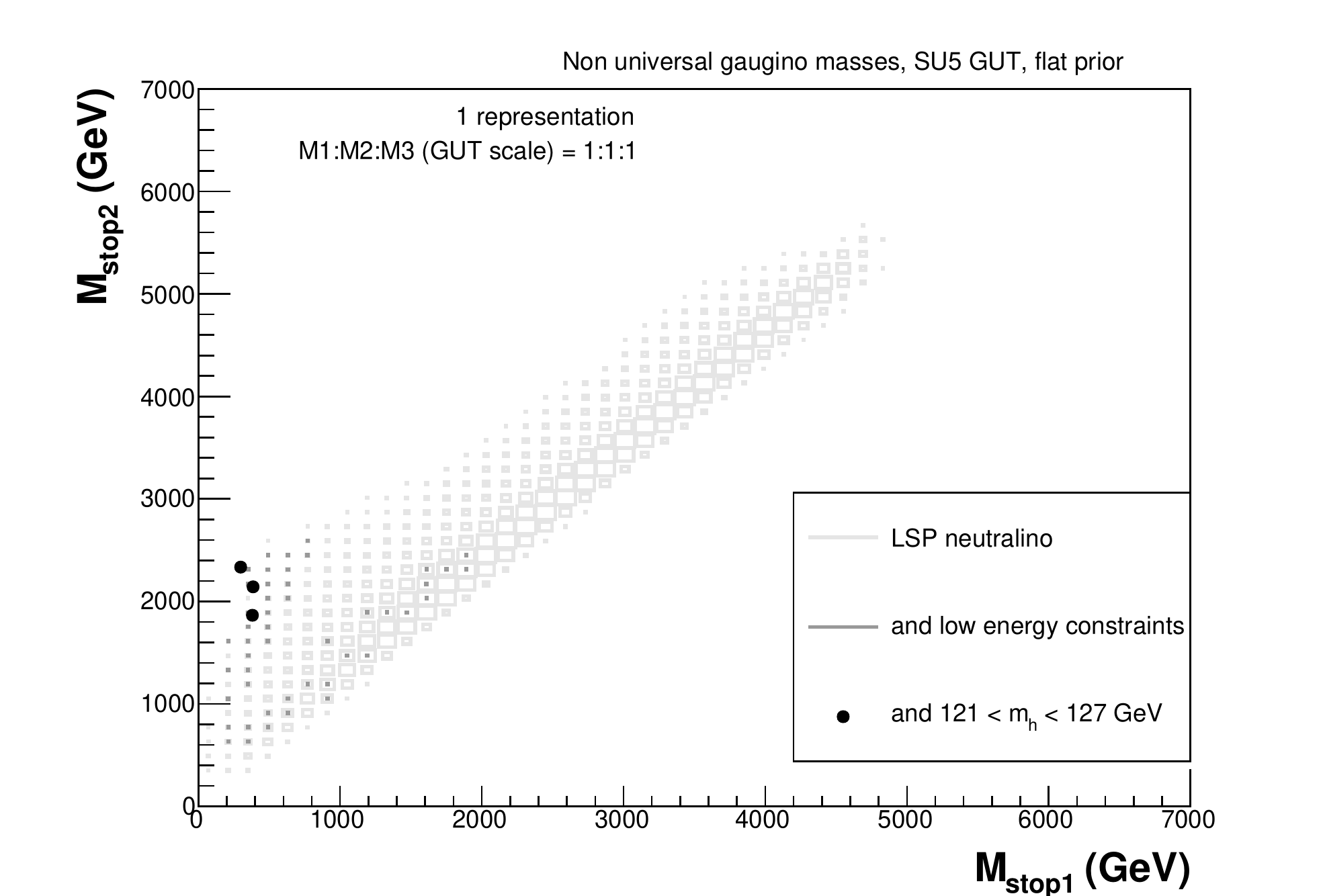}
\hfill
\includegraphics[height=0.49\linewidth,angle=90]{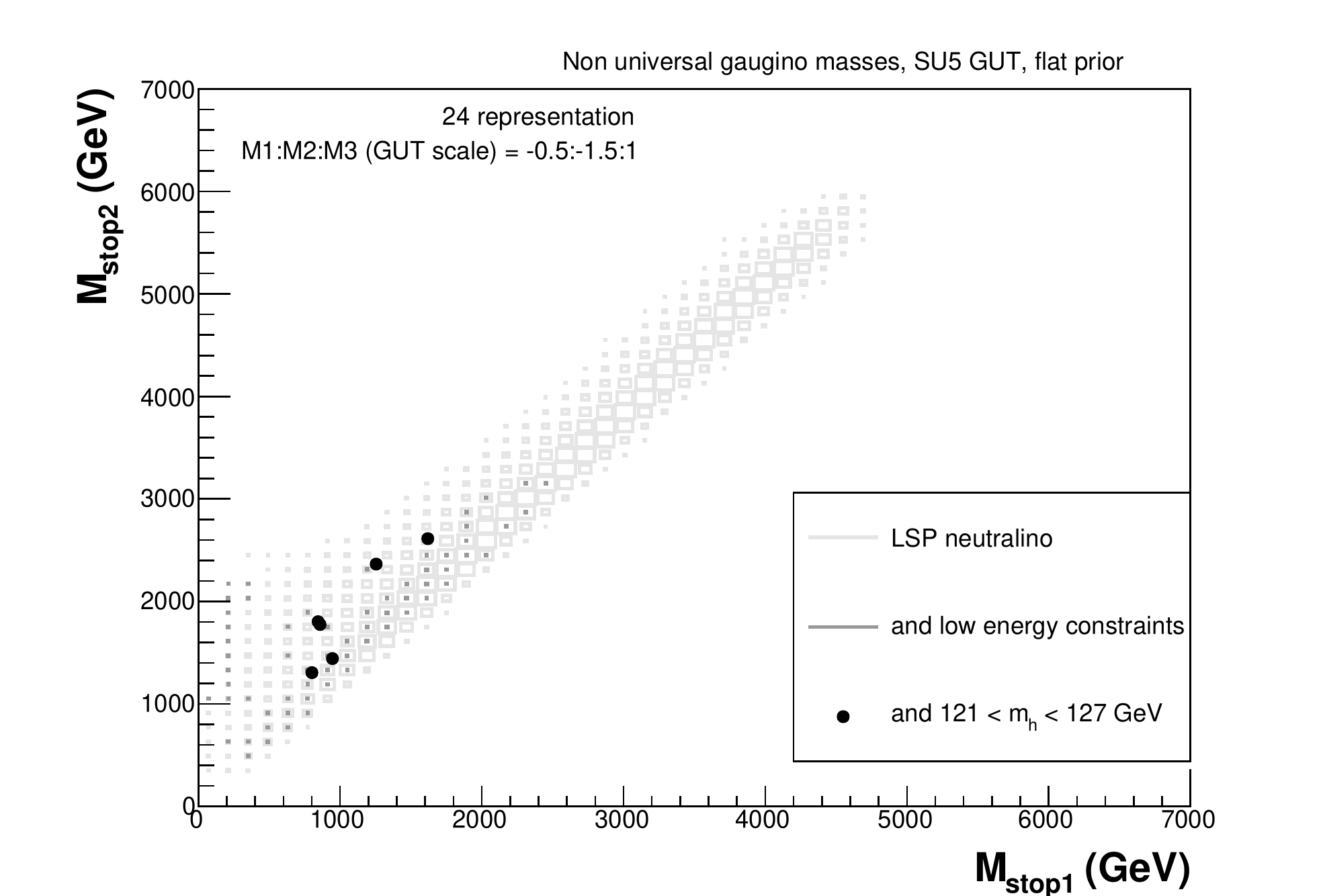}
\includegraphics[height=0.49\linewidth,angle=90]{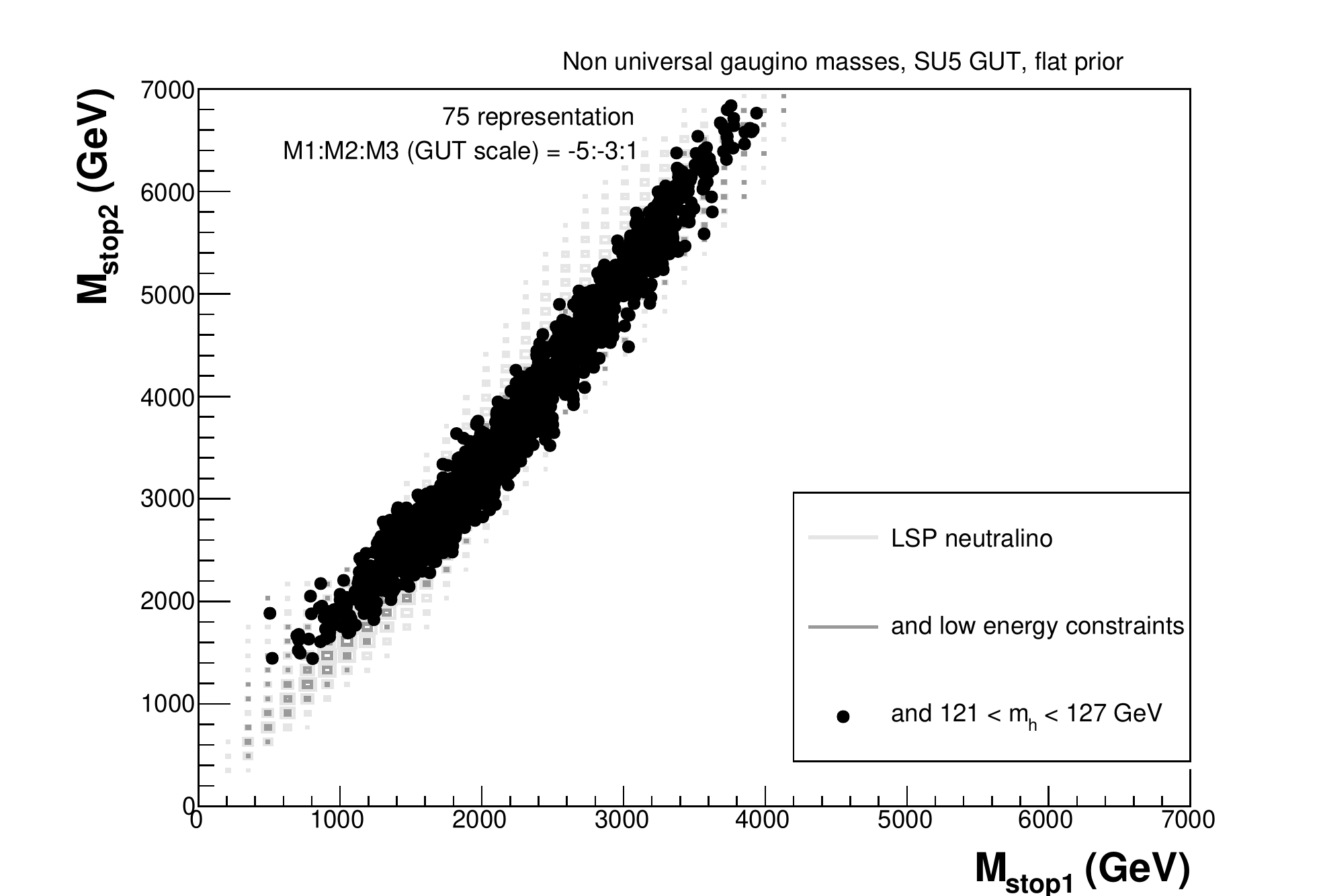}
\hfill
\includegraphics[height=0.49\linewidth,angle=90]{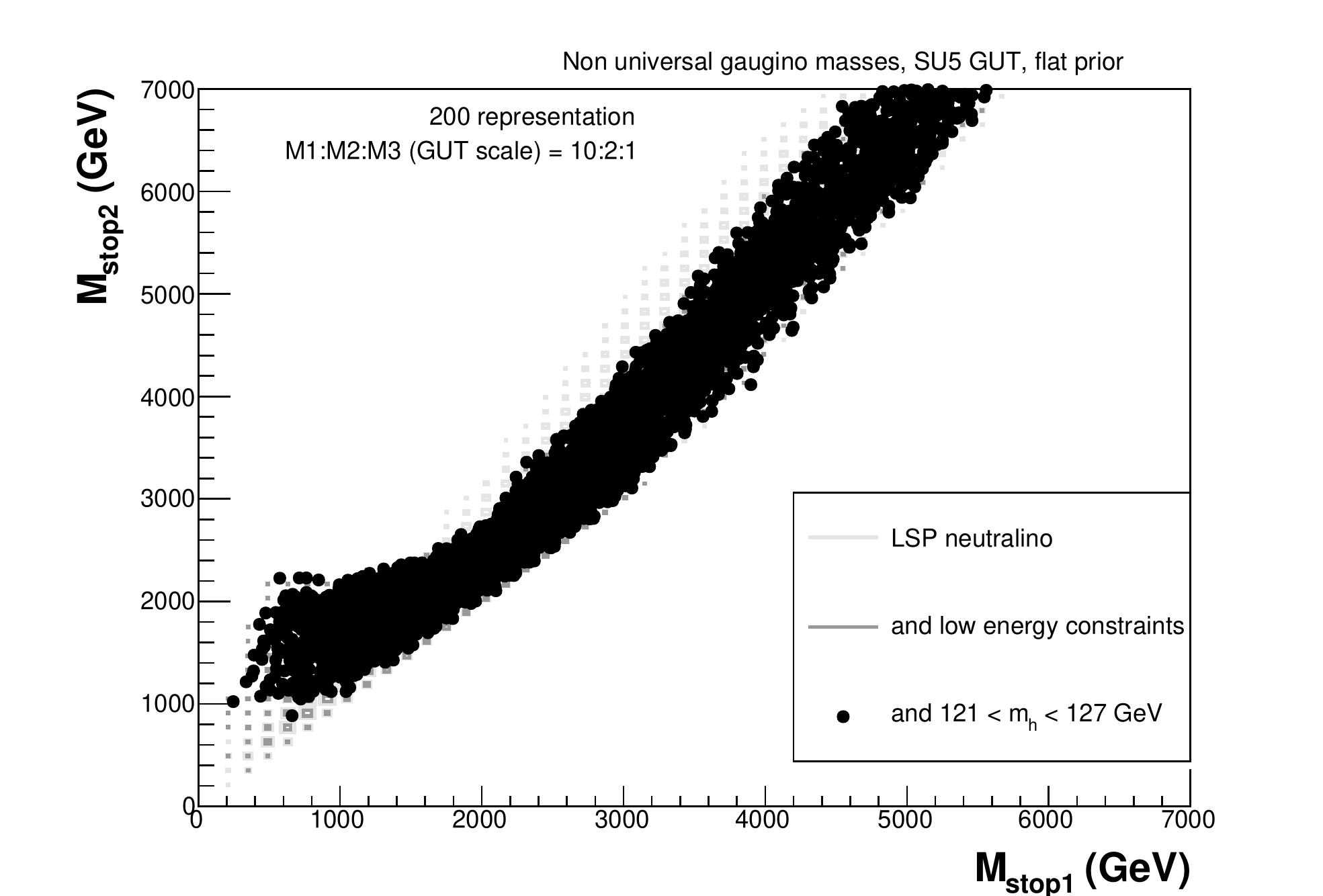}
\caption{Stop masses in the different  representations for the flat
  prior.\label{fig:flatstst}}
\end{figure}

Figure \ref{fig:flatstst} shows the distribution of the two stop
masses for the different representations. By definition, the $\tilde
t_1$ is lighter than the $\tilde t_2$, resulting in the lower bound in
the plot. The points near the upper left edge of the plots have a large
difference between the two stop masses and thus have the strongest
stop mixing.
%%%

For the $\bf 1$ and $\bf24$ representations, the constraints from
section~\ref{sec:allconstraints} conspire to exclude the largest stop
masses. These representations generally have a Bino-like LSP, which
tends to yield a too high DM relic density. After applying the DM
constraint, the only surviving points with high stop masses have
a large $\tan\beta$ value and are excluded by the \Btn constraint. As expected, a relatively high Higgs mass
excludes a scenario where both stops are light. However, the DM constraints favour light scalars to accommodate coannihilation. Indeed, after
implementing the Higgs limits, the surviving points are in the stop and stau coannihilation region, where the stops or the staus are quite heavily mixed.

The situation is quite different for the $\bf75$ and $\bf200$
dimensional representations.  In this case, the large values of $M_1$ and
$M_2$ (cf.~Table~\ref{tab1}) naturally result in higher stop masses after renormalization-group running.
In addition, the DM relic density is less constraining for Wino and
Higgsino-like LSPs, so the constraints from
section~\ref{sec:allconstraints} still allow for large stop
masses. As expected, the Higgs mass excludes both stops being light for these representations as well. This is a fairly general feature of SUSY models, which is not specific to non-universal gaugino mass models.

Summarizing the conclusions we can draw from Figure~\ref{fig:flatstst}, we see for all representations that a relatively heavy Higgs excludes
the lowest stop masses, although the lightest stop can still be well
below 1~TeV due to strong stop mixing.

\begin{figure}[htb]
\includegraphics[height=0.49\linewidth,angle=90]{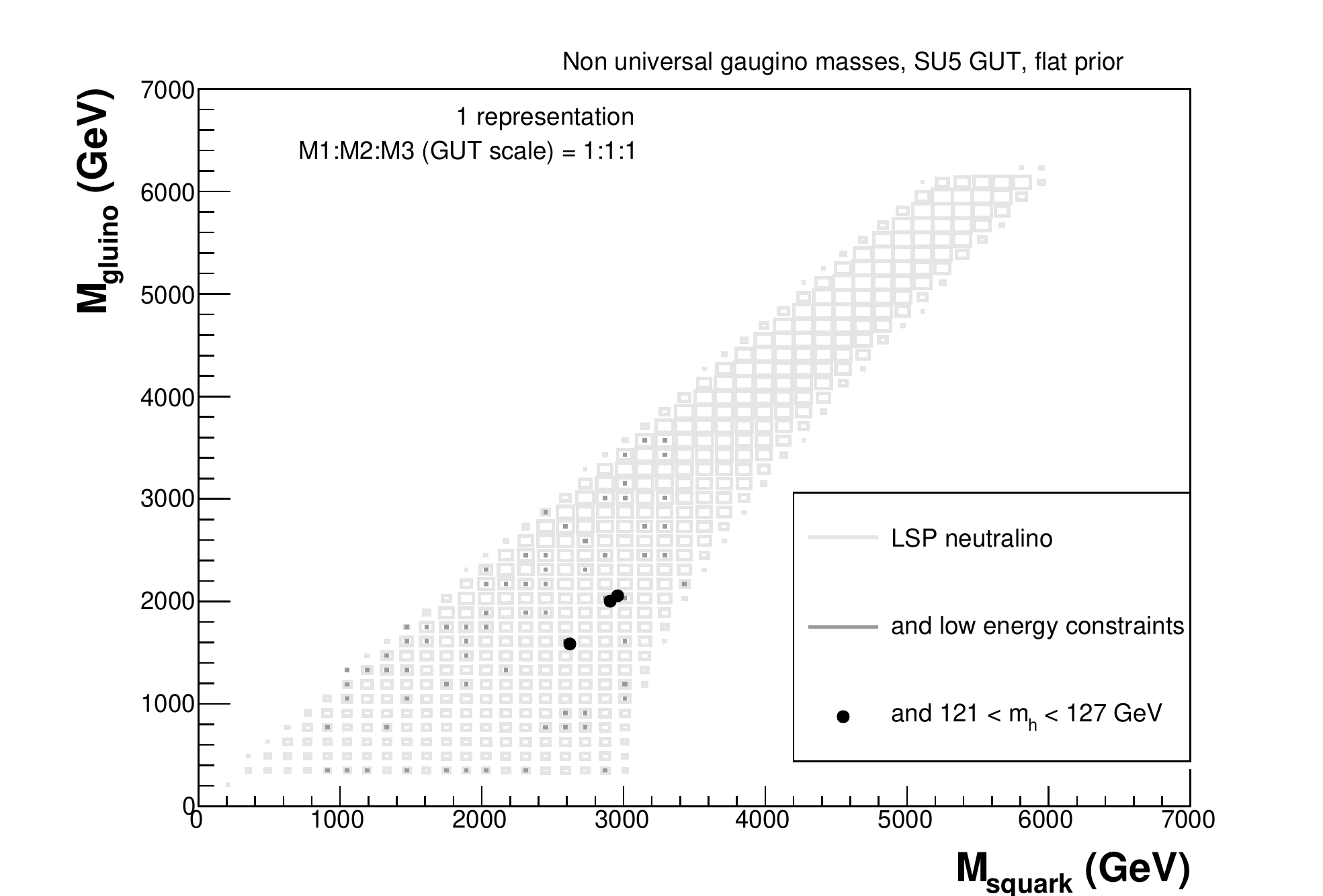}
\hfill
\includegraphics[height=0.49\linewidth,angle=90]{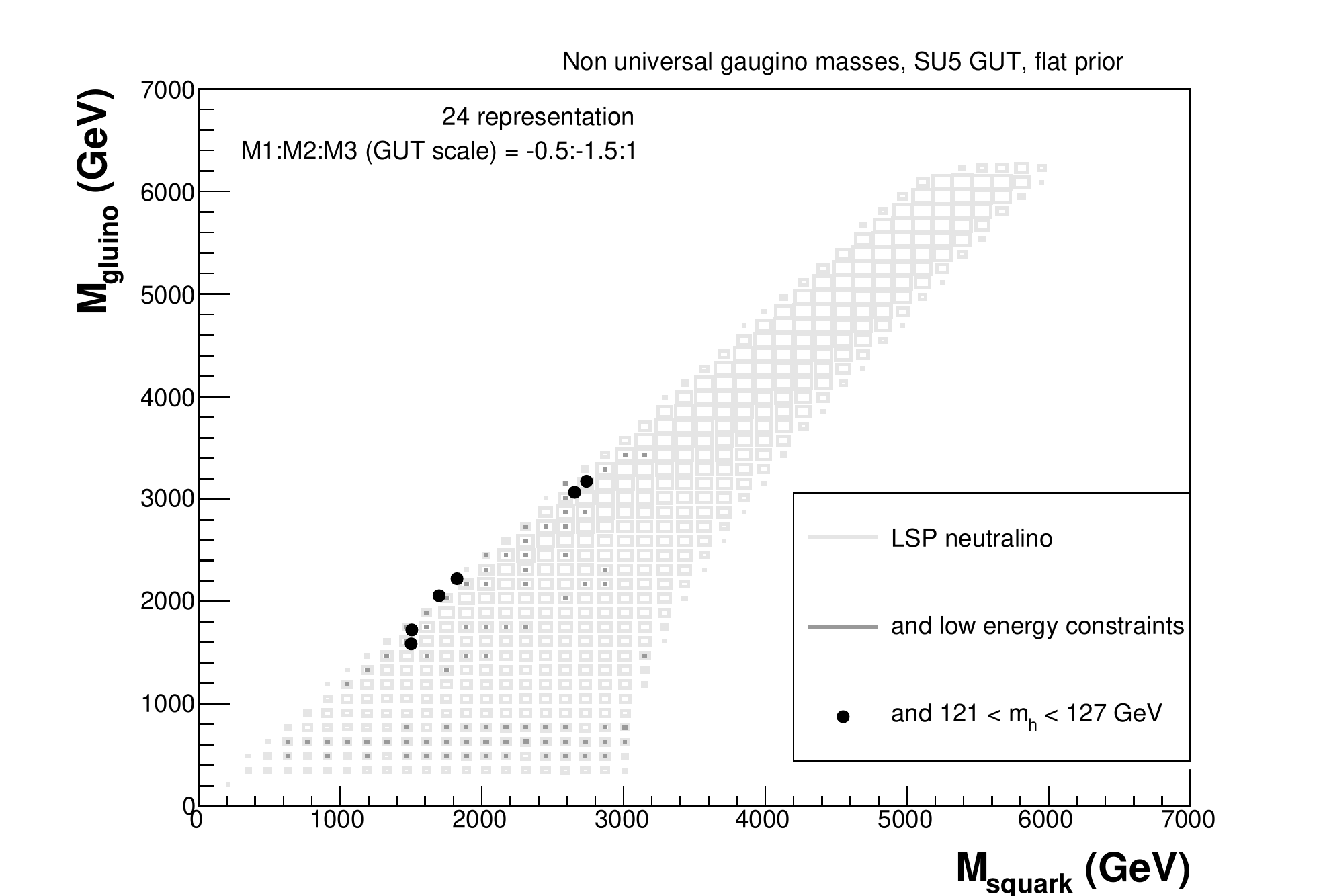}
\includegraphics[height=0.49\linewidth,angle=90]{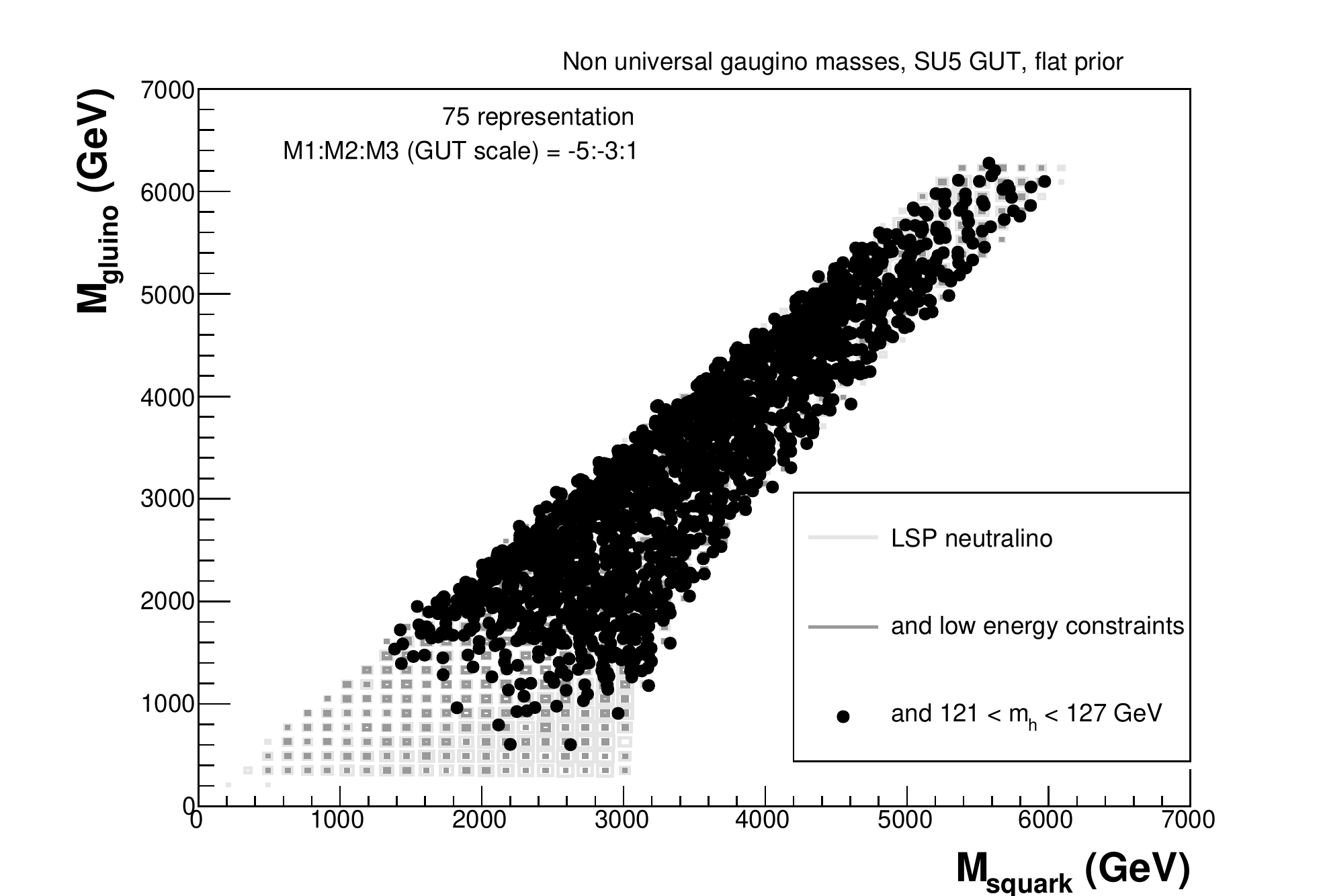}
\hfill
\includegraphics[height=0.49\linewidth,angle=90]{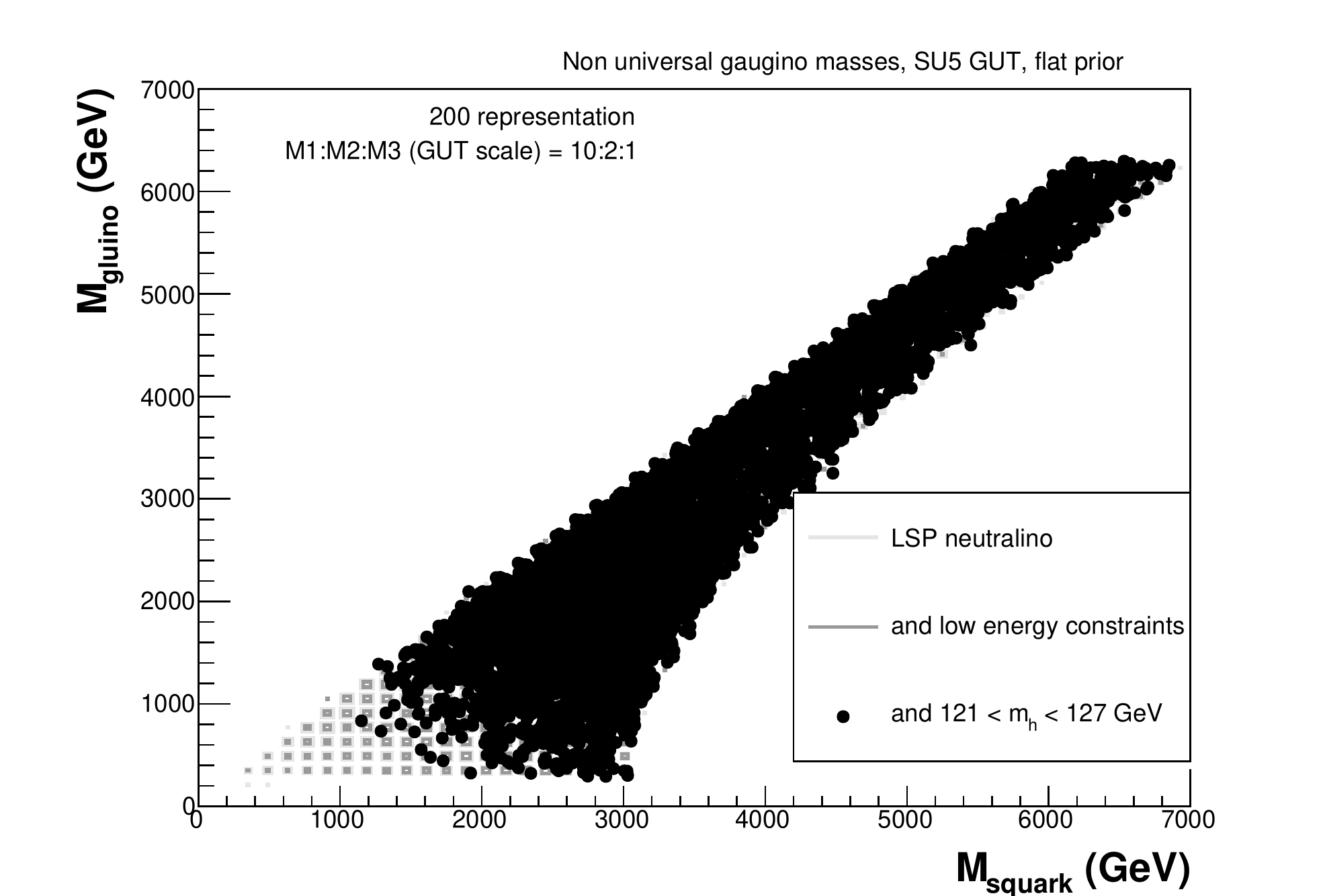}
\caption{Squark and gluino masses in the different representations for the flat prior.\label{fig:flatsqgl}}
\end{figure}

Figure~\ref{fig:flatsqgl} shows the masses of the lightest
light-flavoured squark and the gluino. The almost diagonal upper bound
in the plots comes from the influence of the gluino mass on the squark
mass renormalization group running. The lower bound of the plot is a
result of our scan range for $m_0$.

We see that after implementing the Higgs constraint, the only
points left have squark masses above 1 TeV. The reason
for this is that the stop mass and the squark mass are linked to each
other through $m_0$. As we saw in Fig.~\ref{fig:flatstst}, our Higgs
range selects relatively heavy stops. This naturally implies that
small squark masses are excluded. In the $\bf1$ and $\bf24$ representations, that also results in a large gluino mass. Thus for this relatively large Higgs mass, squarks and gluinos would not have been detected at the LHC yet, a conclusion that was also reached for the $\bf 24$ representation in the context of SO(10) unification \cite{Badziak:2011bn}.

For the $\bf 75$ and $\bf 200$ representations, the Higgs constraint
also tends to prefer higher squark masses due to the need for a higher
stop mass. This kind of squark spectrum correlation is a general feature of models with universal scalar masses. However, due to the different ratios of gaugino masses and their effect on the renormalization-group running, we still have some points left with
low gluino masses. Such points are very interesting in the context of
direct SUSY searches. In the next section, we will investigate whether they could be detected by the LHC.

\section{LHC SUSY Searches}
\label{sec:lhc-susy-searches}

In this section we study the possibility that the low mass SUSY points found in the
previous section are already excluded by the LHC experiments. The mass limits obtained by the LHC experiments, are either model-dependent, or quite weak. In order to assess which surviving points would have been detected by the LHC, we have run the models with either a first or second generation squark or a gluino mass
below 1300 GeV through an emulation of the jet + missing $E_T$ SUSY searches in ATLAS
 \cite{Collaboration:2011qk,Aad:2011ib}. 
 
Events have been generated for each SUSY model with
PYTHIA 6.4 \cite{pythia}. We use DELPHES 1.9 \cite{delphes} as a fast detector
simulation with the default ATLAS detector card, modified by setting the jet cone radius to $0.4$, which is the value used in the ATLAS SUSY searches.
The ATLAS analysis is implemented as in Ref.~\cite{Antonia}, where it is shown that there is good agreement between the ATLAS and DELPHES
setup. Since we are only making an estimate of the exclusion potential, we have not included theoretical uncertainties, even though they would lead to somewhat weaker limits. The event numbers are scaled to the integrated luminosity of the ATLAS experiment with NLO cross sections calculated using PROSPINO2.1~\cite{Beenakker:1996ch,Beenakker:1997ut,prospino}.
These numbers are then used to calculate the expected number of signal events for each signal region and analysis. The results are compared to the
model-independent $95\%$ C.L. limits provided by ATLAS.
Models yielding more events in one of the signal regions are called `excluded'. Models yielding less
events in all signal regions are called `non-excluded'.

\begin{figure}[htb]
\includegraphics[width=0.49\linewidth]{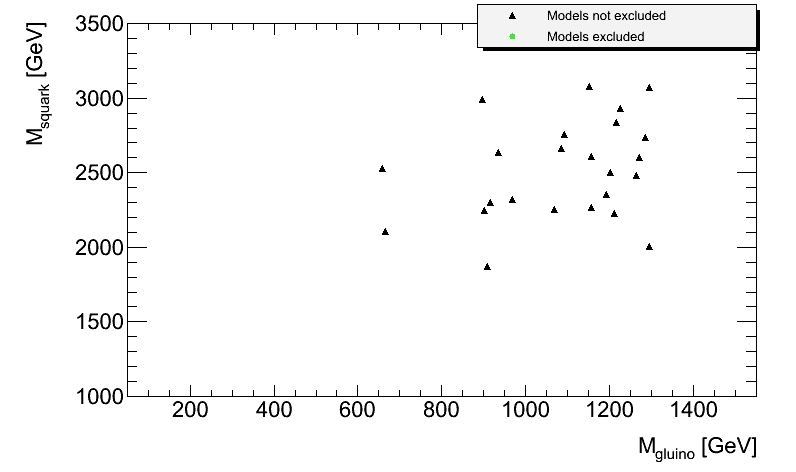}
\hfill
\includegraphics[width=0.49\linewidth]{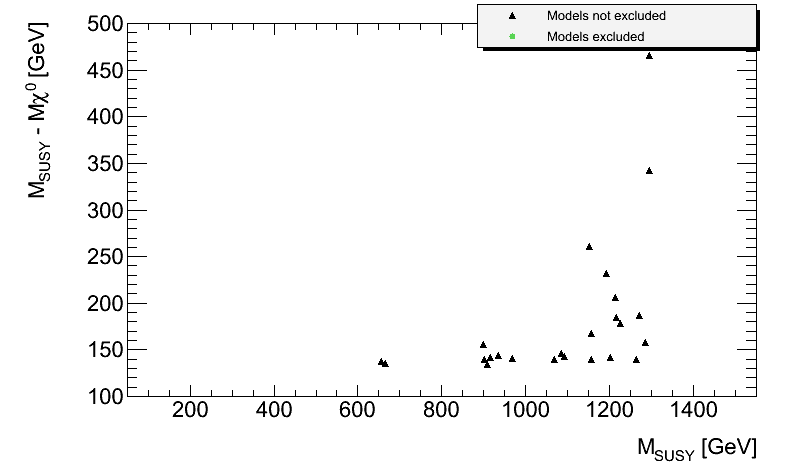}
\includegraphics[width=0.49\linewidth]{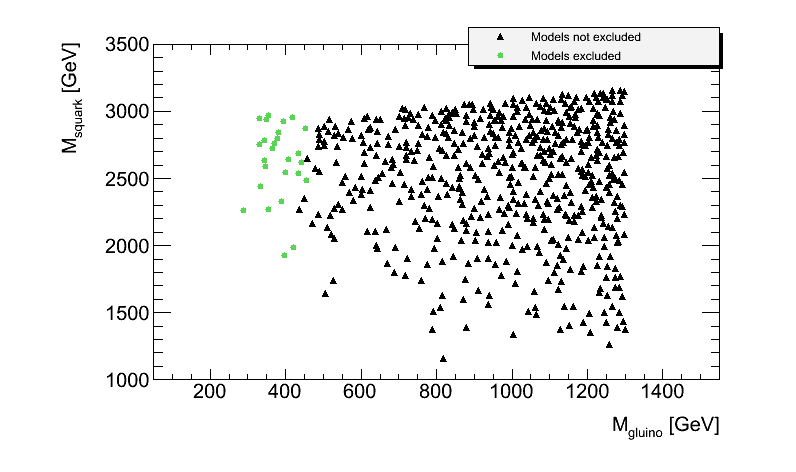}
\hfill
\includegraphics[width=0.49\linewidth]{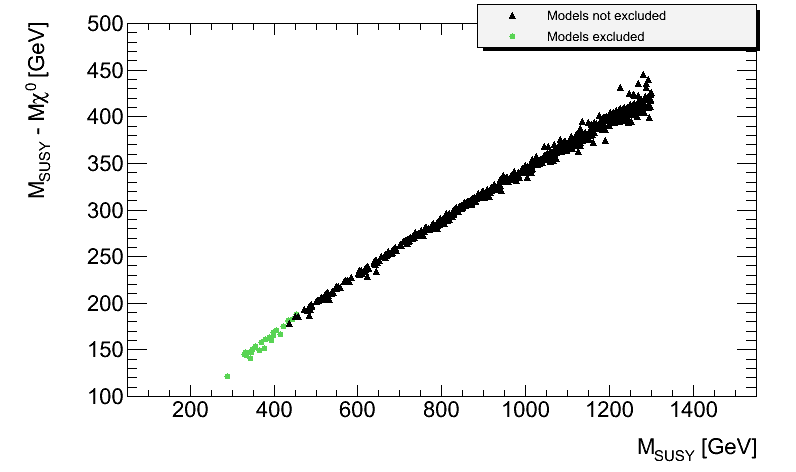}
\caption{The effect of the LHC searches on the surviving low-mass points. The representation {\bf 75} is shown in the top row, while the {\bf 200} representation is shown in the bottom row.\label{fig:atlas}}
\end{figure}

Figure \ref{fig:atlas} shows the results of this study in the squark-gluino mass plane on the left-hand side. On the right-hand side, the same points are shown as a function of $M_{SUSY}$, which is defined as the lightest of the gluino and the light-flavoured squarks, and the mass splitting between this particle and the neutralino. The $\bf75$ representation is shown in the top row and the $\bf200$ representation in the lower row. The black points are not excluded by the LHC, while the green points are.

The shape of gluino-LSP mass splitting is quite different for the two representations. This is caused by the composition of the LSP. In the $\bf75$ representation, the points with $M_\text{SUSY}-m_{\Neu 1} \leq 160$ GeV correspond to a Bino-type neutralino, while the other points correspond to a Higgsino-type neutralino. In the {\bf 200}
representation, all the low-mass points correspond to Wino-type neutralinos.

We see that no models are excluded for the $\bf75$ representation. For the $\bf200$ representation,
models with gluino masses below 400 GeV are excluded by the ATLAS searches. Models with gluino masses
above 400 GeV are not excluded. 
The reason that such low gluino masses are not excluded in these representations is that the mass splittings in these models are small. As can be seen in Figure~\ref{fig:atlas}, the neutralinos tend to be quite heavy compared to the gluino mass. One would need a dedicated search for gluinos with a small or moderate mass difference with the neutralino to exclude these models at the LHC. For part of the model points, the gluino can decay to a stop and a top. In some cases, the subsequent stop decays can produce a total of four tops, leading to spectacular events in the detector. 

Note that the gluino can still escape detection if the mass difference is as large as several hundred GeV. Thus, this problem is not limited to extremely fine-tuned scenarios, but would occur in many models with non-unified masses. A dedicated search for relatively small mass splittings could therefore be useful for a much wider range of SUSY models.

\section{Direct Detection}
\label{sec:direct-detection}

Several direct detection experiments aim at measuring the recoil energy of the nuclei from an elastic
dark matter -- nucleus collision. At present, the most stringent
limits to the spin independent elastic cross section for high weakly-interacting massive particle (WIMP) masses come from the XENON100
experiment \cite{Aprile:2011hi}.
% CDMS  \cite{Ahmed:2009zw}
To compare the calculated proton/neutron cross sections to the
experimental limits, we use a normalized cross section for a
point-like nucleus \cite{Belanger:2010cd}:
\begin{equation}
  \label{eq:SIxs}
  \sigma_\text{SI} = \frac{(Z \sqrt{\sigma^{p}_\text{SI}} +
    (A-Z)\sqrt{\sigma^n_\text{SI}})^2}{A^2},
\end{equation}
$Z$ and $A$ being the atomic and mass number of the target
element.
% (For Xenon: $A=131,\ Z=54$.)
Because there are large uncertainties in the local density of dark
matter and in the nuclear matrix elements that enter the computation
of $\sigma_\text{SI}$ \cite{Ellis:2005mb,Belanger:2008sj}, the direct
detection limits are only indicative.   

\begin{figure}[htb]
\includegraphics[height=0.49\linewidth,angle=90]{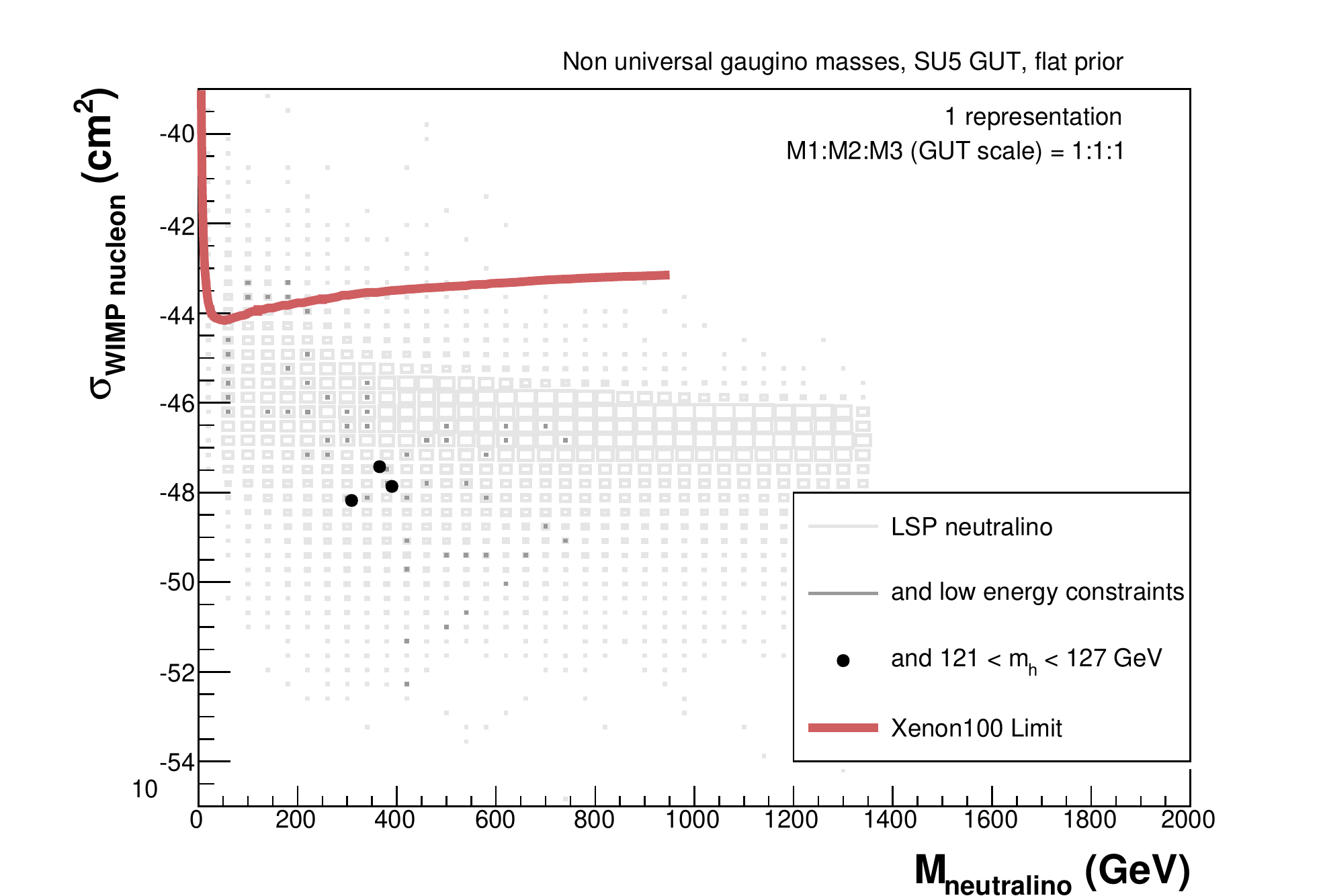}
\hfill
\includegraphics[height=0.49\linewidth,angle=90]{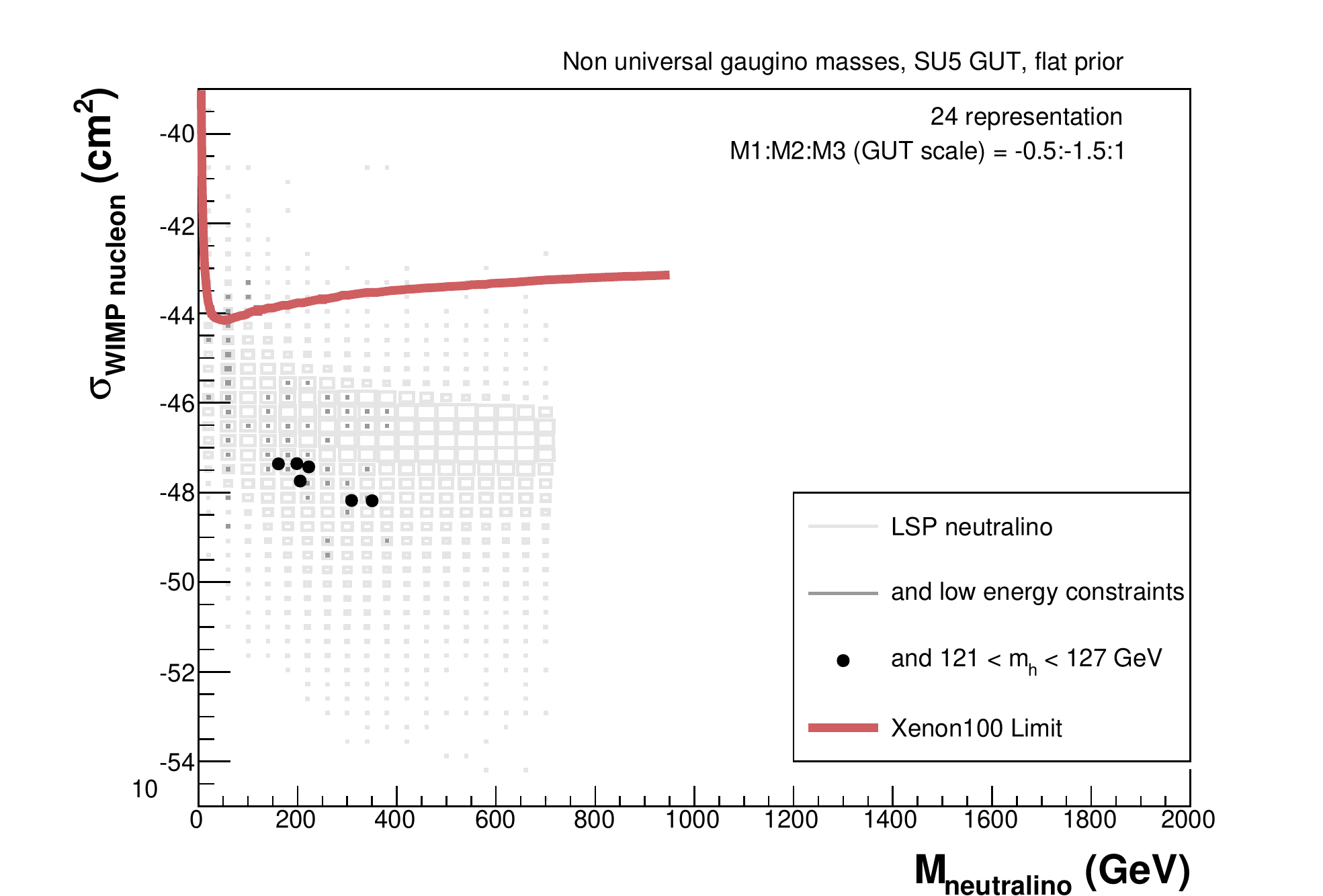}
\includegraphics[height=0.49\linewidth,angle=90]{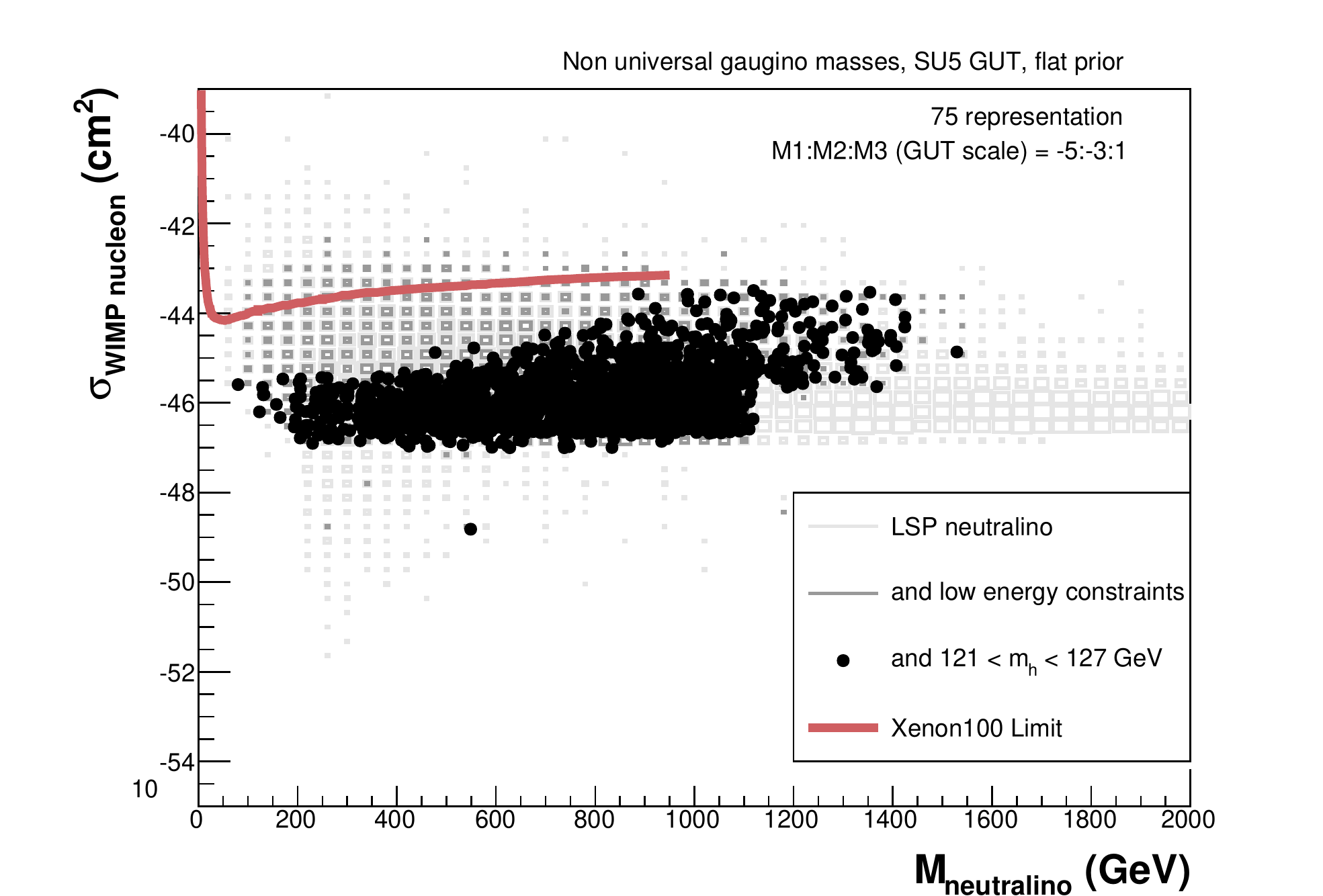}
\hfill
\includegraphics[height=0.49\linewidth,angle=90]{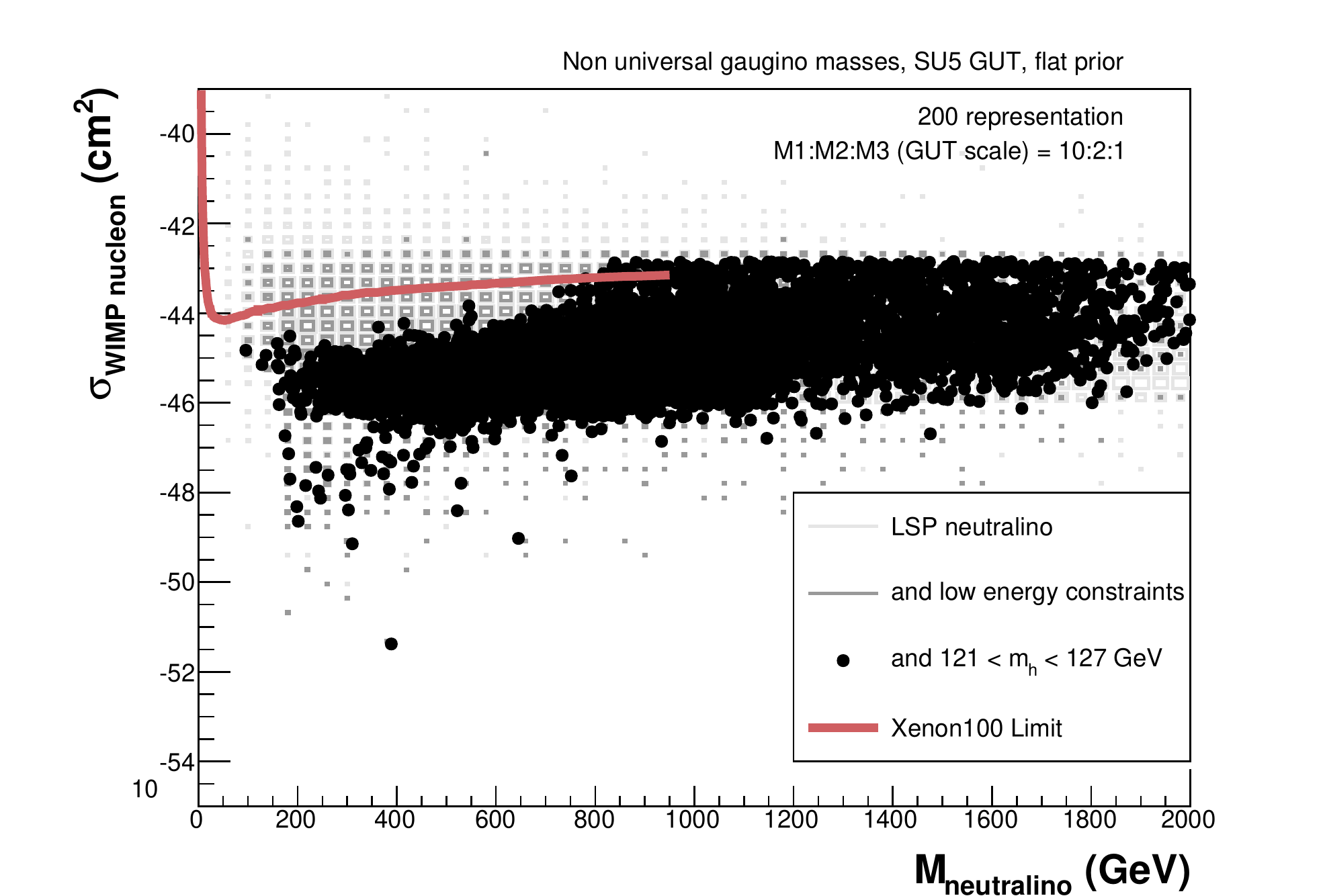}
\caption{Spin-independent WIMP--nucleon cross section for the
  different representations for the flat prior.\label{fig:flatXenon}}
\end{figure}

Figure \ref{fig:flatXenon} shows the neutralino mass and the spin independent cross section for the different representations as well as the XENON100 limit. Once again, the light-grey boxes indicate how many models have a neutralino LSP, the dark grey boxes show the number of models that survive the constraints from section~\ref{sec:allconstraints} and the black points correspond to models that have the correct Higgs mass as well.

We see that most of the models that survive all the constraints from section~\ref{sec:allconstraints} and have the correct Higgs mass cannot be detected by the current direct DM experiments. Only for the {\bf 200}
representation, some of the otherwise allowed points could be excluded
by the XENON100 limit. For the other representations, the surviving
points are well below the direct detection limits, and the
corresponding models are in accordance with all current experimental constraints.

Although the neutralino-nucleon cross section heavily depends on the specific SUSY parameters, the next generation of experiments would be able to detect a large part of the surviving parameter space from the two higher representations. Also, since high neutralino masses are a viable option for Wino and Higgsino LSPs, it would be interesting to extend the XENON100 limit to masses beyond 1 TeV.

\section{Summary and Discussion}
\label{summary}

We have studied the consequences of the recent signs of a Higgs with a mass of 121 -- 127 GeV for SU(5) GUT models with non-universal gaugino masses. We find that results for the different representations can be quite different. We can, however, draw some general conclusions. Models where both stops are light are excluded by the Higgs constraint. As a result, squarks with masses below 1 TeV are excluded as well due to their intimiate connection to stops in models with unified scalar masses. Light gluinos, however, are still an option. These could arise from several models without gaugino mass unification.

We have studied the surviving models with low gluino masses in more detail to see if they would have been observed by the ATLAS direct SUSY searches. We find that indeed some models are excluded. However, for the $\bf 200$ representation, even gluinos with masses below 500~GeV can escape detection due to the small gluino-LSP mass splitting. To detect such models at the LHC, one would most likely need a dedicated search for models with small or moderate mass splittings between the SUSY particles. Such mass splittings could arise in many SUSY scenarios and are not specific for non-universal gaugino mass models.

Finally, we have studied the effect of the Higgs constraint on direct DM detection. We find that if the Higgs is heavy, the current exclusion limits are not sensitive to the largest part of the parameter space. Future experiments would be able to detect many of the interesting points. Also, it would be interesting to extend the range of these searches to higher neutralino masses.

We can conclude that the signs of a heavy Higgs point towards a heavy SUSY mass scale, particularly in the scalar sector. More importantly, most of the points that survive the low-energy, DM and Higgs constraints would not have been detected by the LHC or direct DM searches. Thus, in light of the Higgs results, the non-observation of SUSY so far is no surprise.

\section{Acknowledgments}
We thank Wim Beenakker for useful discussions.
The work of JL and IN was supported by the Foundation for Fundamental
Research of Matter (FOM), program 104 ``Theoretical Particle Physics in
the Era of the LHC".

\bibliographystyle{JHEP}
\bibliography{su5bibs}

\end{document}